\documentclass[conference]{IEEEtran}
\IEEEoverridecommandlockouts
\usepackage{cite}
\usepackage{amsmath,amssymb,amsfonts}
\usepackage[ruled,norelsize,vlined,linesnumbered]{algorithm2e}
\usepackage{multirow}
\usepackage{graphicx}
\usepackage{textcomp}
\usepackage[dvipsnames]{xcolor}

\usepackage{url}
\usepackage{cite}
\usepackage{hyperref}
\usepackage{array}
\usepackage{textcomp}
\usepackage{stfloats}
\usepackage{verbatim}
\usepackage{subfigure}

\usepackage{listings}
\def\BibTeX{{\rm B\kern-.05em{\sc i\kern-.025em b}\kern-.08em
    T\kern-.1667em\lower.7ex\hbox{E}\kern-.125emX}}
\begin{document}
	
\title{HLSPilot: LLM-based High-Level Synthesis\\
\vspace{-0.2em}
}


\makeatletter
\newcommand{\linebreakand}{%
\end{@IEEEauthorhalign}
\hfill\mbox{}\par
\mbox{}\hfill\begin{@IEEEauthorhalign}
}
\makeatother

\author{
	\IEEEauthorblockN{Chenwei Xiong$^{1,2}$, Cheng Liu$^{1,2}$\IEEEauthorrefmark{1}\thanks{\IEEEauthorrefmark{1} Corresponding author.}, Huawei Li$^{1,2}$, Xiaowei Li$^{1,2}$}
	\IEEEauthorblockA{
		$^{1}$SKLP, Institute of Computing Technology, Chinese Academy of Sciences, Beijing, China
	}
	\IEEEauthorblockA{
		$^{2}$Dept. of Computer Science, University of Chinese Academy of Sciences, Beijing, China
	}
	\IEEEauthorblockA{\{xiongchenwei22s, liucheng\}@ict.ac.cn}

 \thanks{This work is supported by the National Key R\&D Program of China under Grant (2022YFB4500405), and the National Natural
Science Foundation of China under Grant 62174162.}
}

\maketitle

\begin{abstract}
Large language models (LLMs) have catalyzed an upsurge in automatic code generation, garnering significant attention for register transfer level (RTL) code generation. Despite the potential of RTL code generation with natural language, it remains error-prone and limited to relatively small modules because of the substantial semantic gap between natural language expressions and hardware design intent. In response to the limitations, we propose a methodology that reduces the semantic gaps by utilizing C/C++ for generating hardware designs via High-Level Synthesis (HLS) tools. Basically, we build a set of C-to-HLS optimization strategies catering to various code patterns, such as nested loops and local arrays. Then, we apply these strategies to sequential C/C++ code through in-context learning, which provides the LLMs with exemplary C/C++ to HLS prompts. With this approach, HLS designs can be generated effectively. Since LLMs still face problems in determining the optimized pragma parameters precisely, we have a design space exploration (DSE) tool integrated for pragma parameter tuning. Furthermore, we also employ profiling tools to pinpoint the performance bottlenecks within a program and selectively convert bottleneck components to HLS code for hardware acceleration. By combining the LLM-based profiling, C/C++ to HLS translation, and DSE, we have established HLSPilot—the first LLM-enabled high-level synthesis framework, which can fully automate the high-level application acceleration on hybrid CPU-FPGA architectures. According to our experiments on real-world application benchmarks, HLSPilot achieve comparable performance in general and can even outperform manually crafted counterparts, thereby underscoring the substantial promise of LLM-assisted hardware designs.

\end{abstract}


\IEEEpeerreviewmaketitle

\section{Introduction}
Hardware designing is a demanding task requiring a high level of expertise. Traditional hardware design involves coding with register transfer level (RTL) language. However, as the complexity of hardware increases continuously with the computing requirements of applications, RTL coding becomes exceedingly time-consuming and labor-intensive. The emergence of High-Level Synthesis (HLS) enables hardware design at higher abstraction levels \cite{martin2009high}. HLS typically employs high-level languages like C/C++ for hardware description, allowing software engineers to also engage in hardware development, which significantly lowering the expertise barrier in hardware design. Designers can focus more on the applications and algorithms rather than the details of low-level hardware implementations. HLS tools automate the design tasks such as concurrent analysis of algorithms, interface design, logic unit mapping, and data management, thereby substantially shortening the hardware design cycle. 

While HLS offers numerous advantages such as higher development efficiency and lower design barriers \cite{martin2009high} \cite{liu2019obfs}, there are still some issues in the real-world HLS-based hardware acceleration workflow \cite{zhang2024graphitron}. Firstly, the overall analysis of the program is of great importance, determining the performance bottlenecks of the program and the co-design between CPU and FPGA remains a challenging issue. Besides, designs based on HLS still encounter a few major performance issues \cite{lahti2018we} \cite{schafer2019high}. Foremost, it still requires substantial optimization experience to craft high-quality HLS code and achieve desired performance in practical development processes \cite{zhao2019performance} \cite{liu2015quickdough}. In addition, HLS code often struggles to reach optimality due to the large design space of various pragma parameters. Some design space exploration (DSE) tools have been proposed \cite{sohrabizadeh2022autodse} \cite{choi2018hls} \cite{zhong2017design} \cite{ferretti2022graph} to automate the parameter tuning, but these tools do not fundamentally optimize the hardware design.
High-quality HLS design turns out to be the major performance challenge from the perspective of general software designers. Some researchers have attempted to address this challenge by using pre-built templates for specific domain applications \cite{luo2023deepburning} \cite{chen2021thundergp} \cite{liang2020deepburning}. For example, ThunderGP \cite{chen2021thundergp} has designed a set of HLS-based templates for optimized graph processing accelerator generation, allowing designers to implement various graph algorithms by filling in the templates. However, it demands comprehensive understanding of both the domain knowledge and the HLS development experience from designers and there is still a lack of well-established universal solution to obtain optimized HLS code. Bridging the gap between C/C++ and HLS remains a formidable challenge requiring further efforts. 

Large Language Models (LLMs) have recently exhibited remarkable capabilities in various generative tasks, including text generation, machine translation, and code generation, underscoring their advanced learning and imitation skills. These advancements have opened up possibilities for addressing hardware design challenges. Researchers have begun applying LLMs to various hardware design tasks, including general-purpose processor designs, domain-specific accelerator designs, and arbitrary RTL code generation. Among these applications, it can be observed that neural network accelerator generation utilizing a predefined template, as reported in \cite{fu2023gpt4aigchip}, reaches an almost 100\% success rate. In contrast, generating register transfer level (RTL) code from natural language descriptions, such as design specifications, experiences a considerably higher failure rate \cite{chang2023chipgpt} \cite{liu2023chipnemo}. This disparity is largely due to the semantic gap between inputs and the anticipated outputs. Despite the imperfections, these work have demonstrated the great potential of exploring LLMs for hardware designing. 

Inspired by prior works, we introduce HLSPilot, an automated framework that utilizes LLMs to generate and optimize HLS code from sequential C/C++ code. Instead of generating RTL code from natural language directly, HLSPilot mainly leverages LLMs to generate the C-like HLS code from C/C++ with much narrower semantic gap and outputs RTL code eventually using established HLS tools. Essentially, HLSPilot accomplishes RTL code generation from C/C++ without imposing hardware design tasks with broad semantic gap on LLMs. Specifically, HLSPilot initiates the process with runtime profiling to pinpoint code segments that are the performance bottleneck and require optimization. Subsequently, HLSPilot extracts the kernel code segments and applies appropriate HLS optimization strategies to the computing kernels to generate optimized HLS code. Then, HLSPilot employs a design space exploration (DSE) tool to fine-tune the parameters of the generated HLS design. Finally, HLSPilot leverages Xilinx OpenCL APIs to offload the compute kernels to the FPGA, facilitating the deployment of the entire algorithm on a hybrid CPU-FPGA architecture. In summary, LLMs are utilized for the hardware acceleration throughout the entire hardware acceleration workflow ranging from profiling, HW/SW partitioning, HLS code generation, HLS code optimization, and tool usage, thereby achieving a high degree of design automation. 


The major contributions of this work are summarized as follows:
\begin{itemize}
	\item We propose HLSPilot, the first automatic HLS code generation and optimization framework from sequential C/C++ code using LLM. This framework investigates the use of LLM for HLS design strategy learning and tool learning, and build a complete hardware acceleration workflow ranging from runtime profiling, kernel identification, automatic HLS code generation, design space exploration, and HW/SW co-design on a hybrid CPU-FPGA computing architecture. The framework is open sourced on Github\footnote{\url{https://github.com/xcw-1010/HLSPilot}}.
 
	\item We propose a retrieval based approach to learn the HLS optimization techniques and examples from Xilinx user manual and utilize an in-context learning approach to apply the learned HLS optimizations on serial C/C++ code and generate optimized HLS code with LLM for various computing kernels.
 
	\item According to our experiments on an HLS benchmark, HLSPilot can generate optimized HLS code from sequential C/C++ code and the resulting designs can outperform manual optimizations with the assistance of DSE tools in most cases. In addition, we also demonstrate the successful use of HLSPilot as a complete hardware acceleration workflow on a hybrid CPU-FPGA architecture with a case study.
\end{itemize}

\section{Related Work}
\subsection{LLM for Hardware Design}
Recent works have begun to utilize LLMs to assist the hardware designing from different angles \cite{thakur2023benchmarking, liu2023verilogeval, tsai2023rtlfixer, liu2023rtlcoder, thakur2023autochip, lu2024rtllm, blocklove2023chip, chang2023chipgpt, fu2023gpt4aigchip, jiang2024iicpilot}. 
Generating RTL code with natural language is a typical approach of hardware design with LLMs. For instance, VGen \cite{thakur2023benchmarking} leverages an open-source LLM, CodeGen \cite{nijkamp2022codegen}, fine-tuned with Verilog code corpus to generate Verilog code. Similarly, VerilogEval \cite{liu2023verilogeval} enhances the LLM's capability to generate Verilog by constructing a supervised fine-tuning dataset, it also establishes a benchmark for evaluating LLM's performance. 
ChipChat \cite{blocklove2023chip} achieves an 8-bit accumulator-based microprocessor design through multi-round natural language conversation. ChipGPT \cite{chang2023chipgpt} proposes a four-stage zero-code logic design framework based on GPT for hardware design.  These studies have successfully applied LLMs to practical hardware designing. However, these methods are mostly limited to small functional modules and the success rate drops substantially when the hardware design gets larger. GPT4AIGchip proposed in \cite{fu2023gpt4aigchip} can also leverage LLMs to generate efficient AI accelerators based on a hardware template, but it relies on pre-built hardware library that requires intensive understanding of both the domain knowledge and the hardware design techniques. which can hinders its use by software developers.
Recently, a domain-specific LLM for chip design, ChipNeMo \cite{liu2023chipnemo}, was proposed. ChipNeMo employs a series of domain-adaptive techniques to train the LLM capable of generating RTL code, writing EDA tool scripts, and summarizing bugs. While powerful, domain-specific LLMs face challenges such as high training costs and difficulties in data collection. 
\subsection{LLM for Code Generation}
Code generation is one of the key applications of LLMs. A number of domain-specific LLMs such as CodeGen \cite{nijkamp2022codegen}, CodeX \cite{chen2021evaluating}, and CodeT5 \cite{wang2021codet5} have been proposed to address the programming of popular languages such as C/C++, Python, and Java, which have a large number of corpus for pre-training and fine-tuning. In contrast, it can be challenging to collect sufficient corpus for the less popular languages. VGen \cite{thakur2023benchmarking} collected and filtered Verilog corpus from Github and textbooks, obtaining only hundreds of MB of corpus. Hence, prompt engineering in combination with in-context learning provides an attractive approach to leverage LLMs to generate code for domain-specific languages. For instance, the authors in \cite{wang2024grammar} augment code generation by providing the language's  Backus–Naur form (BNF) grammar within prompts. 

\section{HLSPilot Framework}

The remarkable achievements of LLMs across a wide domain of applications inspire us to create an LLM-driven automatic hardware acceleration design framework tailored for a hybrid CPU-FPGA architecture. Unlike previous efforts that primarily focused on code generation, our objective is to harness the potential of LLMs to emulate the role of an expert engineer in hardware acceleration. Given that hardware acceleration on a hybrid CPU-FPGA architecture demands a set of different design tasks such as runtime profiling, compute kernel identification, compute kernel acceleration, design space exploration, and CPU-FPGA co-design, LLMs must understand the design guidelines and manipulate the relevant design tools to achieve the desired design objectives, akin to an engineer. Fortunately, LLMs have exhibited powerful capabilities in document comprehension, in-context learning, tool learning, and code generation, all of which align perfectly with the hardware acceleration design requirements. The intended design framework eventually provides an end-to-end high-level synthesis of sequential C/C++ code on a hybrid CPU-FPGA architecture, thus named as HLSPilot, which will be elaborated in the rest of this section.  

\subsection{HLSPilot Overview}

\begin{figure*}[htp]
    \centering
    \includegraphics[width=\textwidth]{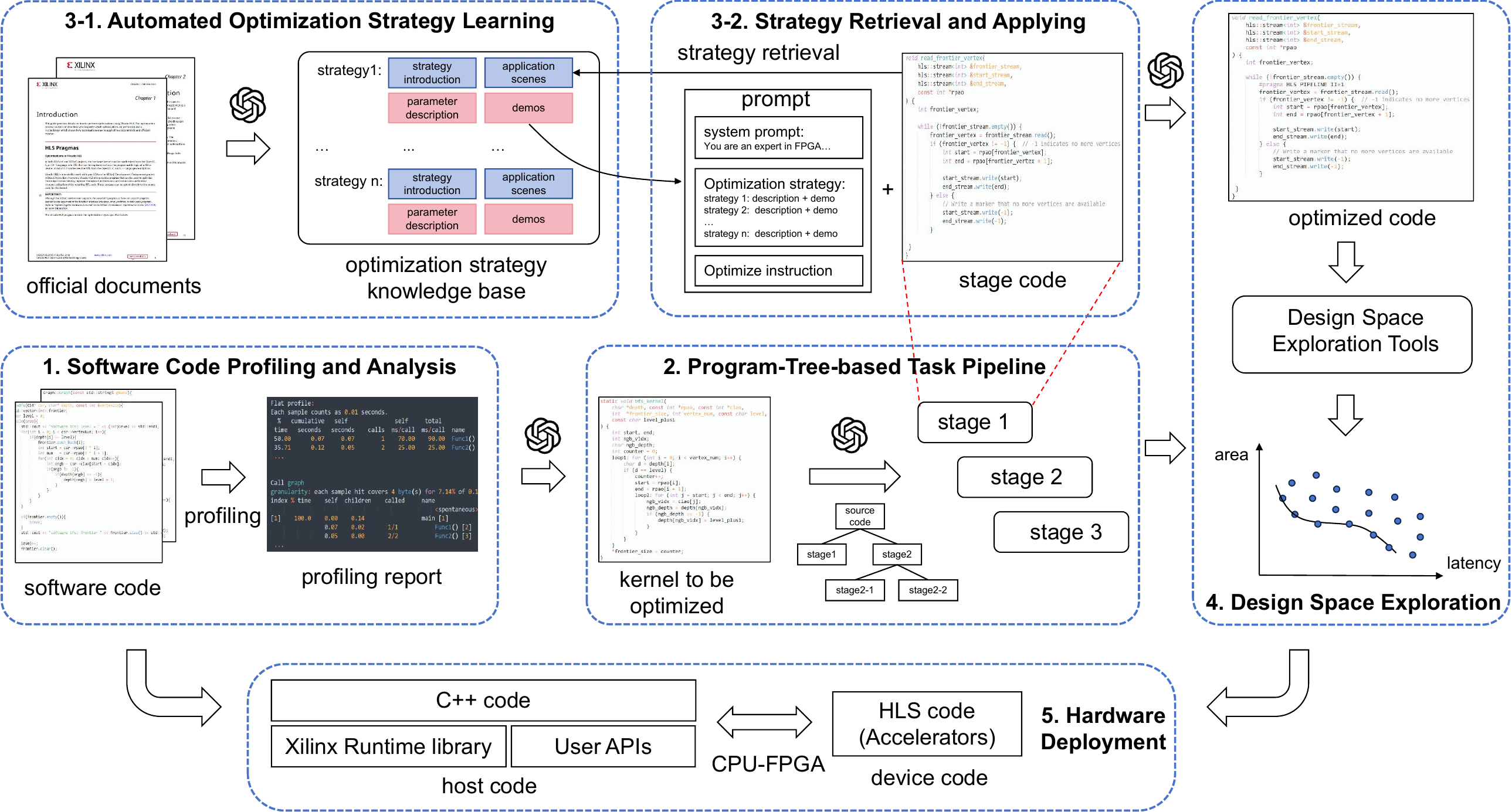}
    \caption{HLSPilot framework}
    \label{fig:overview}
    \vspace{-1.2em}
\end{figure*} 

HLSPilot as presented in Fig. \ref{fig:overview} takes sequential C/C++ code as design input and it mainly includes five major processing stages to generate optimized hardware acceleration solution on a hybrid CPU-FPGA architecture. 

Firstly, HLSPilot conducts runtime profiling on the high-level application code to identify the most time-consuming computing kernels, which will be the focus of subsequent optimization. In this work, we profile the target algorithm and analyze the execution time with gprof on a CPU system. Then, a detailed performance report will be generated as needed. With the report, we can obtain the performance information such as execution time distribution across the algorithm and the number of function calls conveniently. Since LLMs is capable to understand and summarize the textual reports, the time-consuming functions can be identified conveniently. HLSPilot extracts the computing kernels to be optimized in next stage based on these profiling information. 

Secondly, the computing kernels are organized as dependent tasks and pipelined accordingly. The dependent tasks can be implemented efficiently with the data flow mechanism supported by Xilinx HLS. While the compute kernels can be irregular, we propose a program-tree-based strategy to refactor the program structure of the compute kernels and generate an optimized task flow graph while ensuring equivalent code functionality. Details of the automatic task pipelining will be illustrated in Section \ref{sec:task-pipeline}.

Thirdly, we start to optimize each task with HLS independently. While there are many distinct HLS optimization strategies applicable to different high-level code patterns, we create a set of HLS optimization strategies based on Xilinx HLS user guide and leverage LLMs to select and apply the appropriate optimization strategies automatically based on the code patterns in each task. Details of the LLM-based automatic HLS optimization will be presented in Section \ref{sec:hls-opt}.

Fourthly, after the code refactoring and the application of various HLS pragmas, the HLS code can be obtained, but the parameters such as the initiation interval (II) for pipelining, the factors of loop unrolling, and the size for array partitioning in the HLS code still needs to be tuned to produce accelerators with higher performance. However, it remains rather challenging for LLMs to decide design parameters of a complex design precisely. To address this issue, HLSPilot utilizes external tools to conduct the design space exploration and decides the optimized solution automatically. According to recent research \cite{schick2024toolformer}, LLMs is capable to learn and utilize external APIs and tools efficiently. Hence, HLSPilot leverages LLMs to extract the parameters from HLS code and invoke the DSE tool proposed in \cite{DSEtool} by generating the corresponding execution scripts.

Finally, when the compute kernels are optimized with HLS, they can be compiled and deployed on FPGAs for hardware acceleration. Nonetheless, these accelerators must be integrated with a host processor to provide a holistic hardware acceleration solution. The acceleration system has both host code and device code that will be executed on CPU side and FPGA side respectively. HLSPilot leverages LLMs to learn the APIs provided by Xilinx runtime (XRT) to manage the FPGA-based accelerators and perform the data transfer between host memory and FPGA device memory. Then, it generates the host code mostly based on the original algorithm code and replaces the compute kernels with the compute APIs that will invoke the FPGA accelerators and the data movement APIs. The device code is mainly the HLS code generated in prior steps. With both the host code and device code, the entire algorithm can be deployed on the hybrid CPU-FPGA architecture.


\subsection{Program-Tree-based Task Pipelining} \label{sec:task-pipeline}

While the compute kernel can be quite complex, it needs to be split into multiple tasks for the sake of potential pipelining or parallel processing, which is critical to the performance of the generated accelerator. However, it is difficult to split the compute kernel appropriately because inappropriate splitting may lead to imbalanced pipelining and low performance. In addition, the splitting usually causes code refactoring, which may produce code with inconsistent functionality and further complicate the problem. To address this problem, we propose a program-tree-based strategy to guide LLM to produce fine-grained task splitting and pipelining.  

\begin{algorithm}
    \label{algo:task-pipeline}
    \SetAlgoLined 
    \caption{Program-tree-based Pipelining Strategy}
    \KwIn{Top-level Function Code C}
    \KwOut{Tasks Collection $T=\{task_1,task_2,\dots,task_n\}$}
	{$T \gets \{C\}$}\\
        \While{$T$ has $task$ that can be further split} {
		{$T_{new} \gets \{\}$}\\
            \For {$task_i \in T$} {
			\eIf{LLM decides to futher split $task_i$} {
				{1.For non-loop blocks: split the code based on the functionality of the statement execution}\\
				{2.For loop blocks: split the code based on the minimum parallelizable loop granularity}\\
				{Add the refactored code to $T_{new}$}\\
			} {
				{Add $task_i$ to $T_{new}$}\\
			}
		}
            {$T \gets T_{new}$}\\
	}
\end{algorithm}

The proposed program-tree based task pipelining strategy is detailed in Algorithm \ref{algo:task-pipeline}. According to the strategy, LLM iteratively decomposes the compute kernel to smaller tasks and form a tree structure eventually. An input compute kernel $C$ is denoted as the root node of the tree. Hence, the initial node set of the tree $T=\{C\}$. Then, LLM decides whether each $task$ in $T$ can be further decomposed based on the complexity of the task code. If a decomposition is confirmed in $task_i$, LLM will perform the code decomposition. The decomposition for non-loop tasks and loop tasks are different and they will be detailed later in this sub section. If the task cannot be further decomposed, the $task_i$ is added to $T_{new}$ directly. 

The major challenge of the program-tree-based task pipelining strategy is the task decomposition metric which depends on the code structures and can vary substantially. As a result, the metric can be difficult to quantify. Instead of using a determined quantitative metric, we leverage LLMs to perform the task decomposition with natural language rules and typical decomposition examples. Specifically, for non-loop code, we have LLM to analyze the semantics of code statements, recognize the purpose of these statements, and group statements performing the same function into a single task. For loop code, the decomposition is primarily based on the smallest loop granularity that can be executed in parallel. We take advantage of the in-context learning capabilities of LLMs and present a few representative decomposition examples to guide the task decomposition for general scenarios. These examples as detailed as follows.



\subsubsection {Each iteration of the loop is considered as a task}
	In the original merge sort loop, each iteration processes all intervals of the same width. Therefore, each iteration can be regarded as a task. For example, $task_i$ merges all intervals with a width equal to $2^i$.
    \begin{lstlisting}
// before:
for (int width = 1; width < SIZE; width = 2 * width) {
    for (int i1 = 0; i1 < SIZE; i1 = i1 + 2 * width) {
        int i2 = i1 + width;
        int i3 = i1 + 2 * width;
        if (i2 >= SIZE) i2 = SIZE;
        if (i3 >= SIZE) i3 = SIZE;
        merge(A, i1, i2, i3, temp);
    }
}

// after:
for (int stage = 1; stage < STAGES - 1; stage++) {
    // merge all equally wide intervals
    merge_intervals(temp[stage - 1], width, temp[stage]);
    width *= 2;
}
    \end{lstlisting}

\subsubsection {The first and second halves of a loop's traversal are each considered as a task}
	In histogram statistics, since the first and second halves of the loop can be executed in parallel, they are considered as two tasks.
    \begin{lstlisting}
// before:
for (int i = 0; i < INPUT_SIZE; i++) {
    val = in[i];
    hist[val] = hist[val] + 1;
}

// after:
for (int i = 0; i < INPUT_SIZE / 2; i++) {
    val = in1[i];
    hist1[val] = hist1[val] + 1;
}
for (int i = 0; i < INPUT_SIZE / 2; i++) {
    val = in2[i];
    hist2[val] = hist2[val] + 1;
}
histogram_reduce(hist1, hist2, hist);       
    \end{lstlisting}
    
\subsubsection {Each level of a loop is considered as a task}
	In BFS algorithm, there are two loops, with the first loop used to find the frontier vertex and read the corresponding rpao data, the second loop used to traverse the neighbors of the frontier vertex, which can be divided into two tasks based on this.
    \begin{lstlisting}
// before:
loop1: for (int i = 0; i < vertex_num; i++) {
    char d = depth[i];
    if (d == level) {
        start = rpao[i];
        end = rpao[i + 1];
        loop2: for (int j = start; j < end; j++) {
            ngb_vidx = ciao[j];
            ngb_depth = depth[ngb_vidx];
            if (ngb_depth == -1) {
                depth[ngb_vidx] = level_plus1;
            }
        }
    }
}

// after: 
void read_frontier_vertex(int *depth, int vertex_num, int level, int *rpao, ...) {
    ...
    for (int i = 0; i < vertex_num; i++) {
        if (depth[i] == level) {
            int start = rpao[i];
            int end   = rpao[i + 1];
            start_stream << start;
            end_stream   << end;
        }
    }
}
void traverse(hls::stream<int>& start_stream, hls::stream<int>& end_stream, ...) {
    ...
    while (!start_stream.empty() && !end_stream.empty()) {
        int start = start_stream.read();
        int end   = end_stream.read();
        for (int j = start; j < end; j++) {
            ngb_vidx = ciao[j];
            ngb_depth = depth[ngb_vidx];
            if (ngb_depth == -1) {
                depth[ngb_vidx] = level_plus1;
            }
        }
    }
}
		\end{lstlisting}	

\subsubsection {Multiple levels of loops are considered as a task}
	In video frame image convolution, there are a total of 4 layers of loops, where loop1 and loop2 are considered as the tasks for reading the pixel, and loop3 and loop4 are the tasks for calculating the convolution.
    \lstset{
		language=C++, 
		linewidth=0.48\textwidth, 
		columns=flexible, 
		basicstyle=\ttfamily \fontsize{8pt}{9pt}\selectfont , 
		breaklines=true,
		frame=single, 
		keywordstyle=\textcolor{Orchid}, 
		commentstyle=\textcolor{Gray}, 
		stringstyle=\textcolor{Orange}, 
		emph={merge, merge_arrays, histogram, histogram_parallel, histogram_map, histogram_reduce, bfs_kernel, read_frontier_vertex, traverse, convolve, read_dataflow, compute_dataflow, empty, read}, 
    	emphstyle=\textcolor{NavyBlue}
    }
    \begin{lstlisting}
// before: 
loop1: for(int line=0; line<img_h; ++line) {
    loop2: for(int pixel=0; pixel<img_w; ++pixel) {
        float sum_r = 0, sum_g = 0, sum_b = 0;
        loop3: for(int m=0; m<coeff_size; ++m) {
            loop4: for(int n=0; n<coeff_size; ++n) {
int ii = line + m - center;
int jj = pixel + n - center;
if(ii >= 0 && ii < img_h && jj >= 0 && jj < img_w) {
    sum_r += in[(ii * img_w) + jj].r * coeff[(m * coeff_size) + n];
    sum_g += in[(ii * img_w) + jj].g * coeff[(m * coeff_size) + n];
    sum_b += in[(ii * img_w) + jj].b * coeff[(m * coeff_size) + n];
}
    ...
}

// after: 
void read_dataflow(hls::stream<RGBPixel>& read_stream, const RGBPixel *in, int img_w, int elements, int half) {
    int pixel = 0;
    while(elements--) {
        read_stream <<  in[pixel++];
    }
    ...
}

void compute_dataflow(hls::stream<RGBPixel>& write_stream, hls::stream<RGBPixel>& read_stream, const float* coefficient, int img_width, int elements, int center) {
static RGBPixel window_mem[COEFFICIENT_SIZE][MAX_WIDTH];
static fixed coef[COEFFICIENT_SIZE * COEFFICIENT_SIZE];
    for(int i  = 0; i < COEFFICIENT_SIZE*COEFFICIENT_SIZE; i++) {
        coef[i] = coefficient[i];
    }
    ...
}
    \end{lstlisting}

\begin{figure}[htp]
    \centering
    \includegraphics[width=0.48\textwidth]{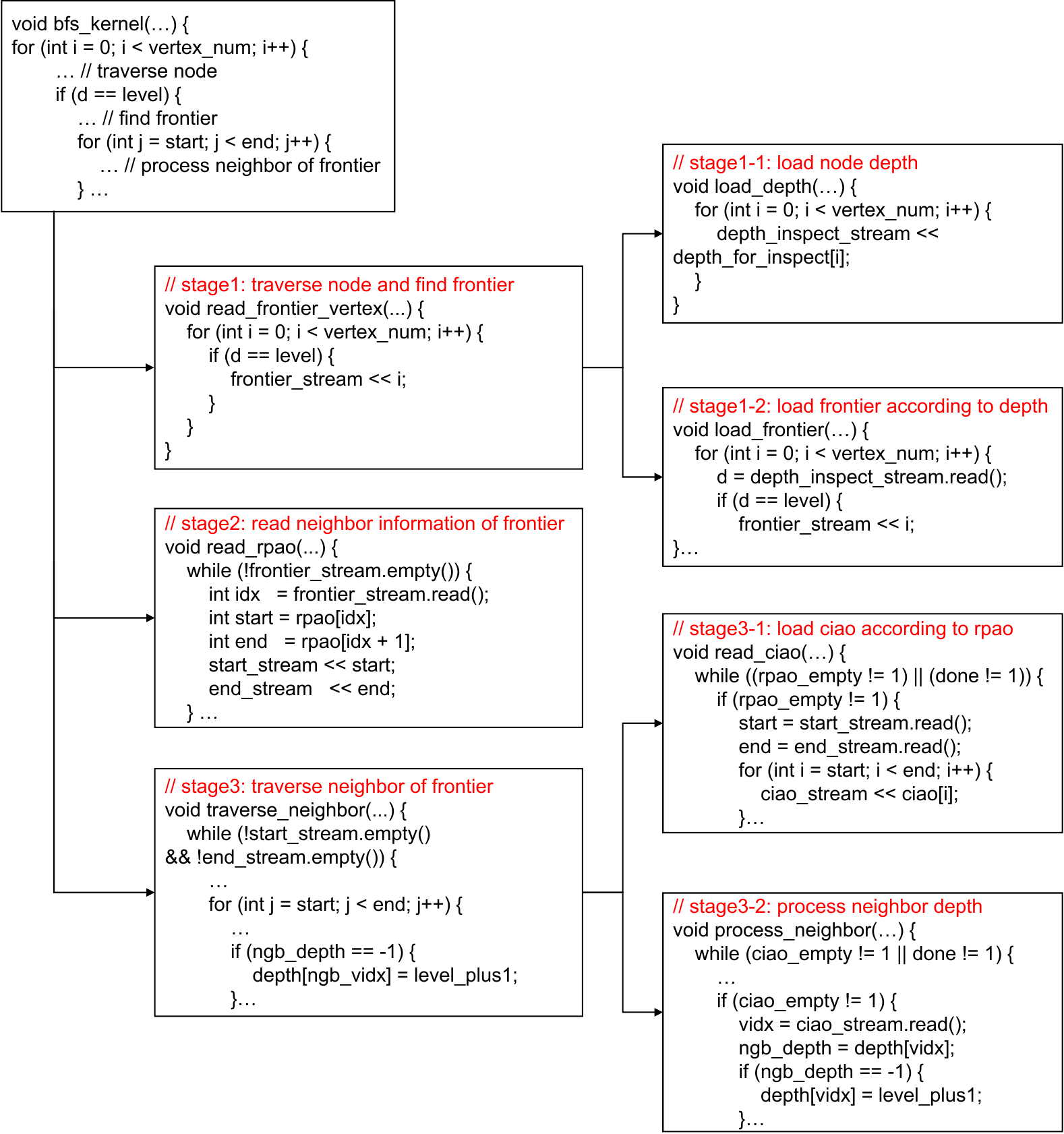}
    \caption{An example of program tree construction. LLM divides BFS with nested loop into multiple dependent tasks for the pipelined execution.}
    \label{fig:program-tree}
    \vspace{-1em}
\end{figure} 

In order to demonstrate the proposed task decomposition strategy, we take BFS with relatively complex nested loop as an example and present the generated program tree in Fig.\ref{fig:program-tree}. It shows that the nested loop in BFS are effectively identified and extracted as dependent tasks correctly.

When the tasks are decomposed, the corresponding code segments will be packed into a function and the code needs to be refactored accordingly. Before proceeding to the HLS acceleration, HLSPilot needs to check the correctness of the refactored code. Specifically, we compare the refactored code to the original code by testing the execution results to ensure the computing results are consistent. We follow a bottom-up testing strategy and start from the leaf nodes of the program tree. If an error occurs, it can be traced back to the erroneous leaf node and check from its parent node. If errors persist across multiple attempts, the program tree is backtracked and the parent node is considered as the final refactored result.  

\subsection{LLM-based Automatic HLS Optimization}\label{sec:hls-opt}
After the task pipelining, we continue to apply appropriate HLS optimization strategies to these tasks. The HLS optimization strategies are mainly extracted from Vendor's documentation \cite{XilinxUG902} \cite{XilinxUG1270} \cite{XilinxUG1399} by LLM. Since the optimizations are usually limited to specific scenarios or code patterns, there are a number of distinct strategies but only a few of them may be actually utilized for a specific compute kernel in practice. 
To facilitate the automatic HLS optimization, we build an HLS optimization strategy knowledge base and propose a Retrieval-Augmented-Generation-like (RAG-like) strategy to select the most suitable optimization strategies from knowledge base. The selected optimization strategies will be applied to the target code through in-context learning, ensuring optimized HLS code generation. 



\begin{figure*}[htp]
    \centering
    \includegraphics[width=0.95\textwidth]{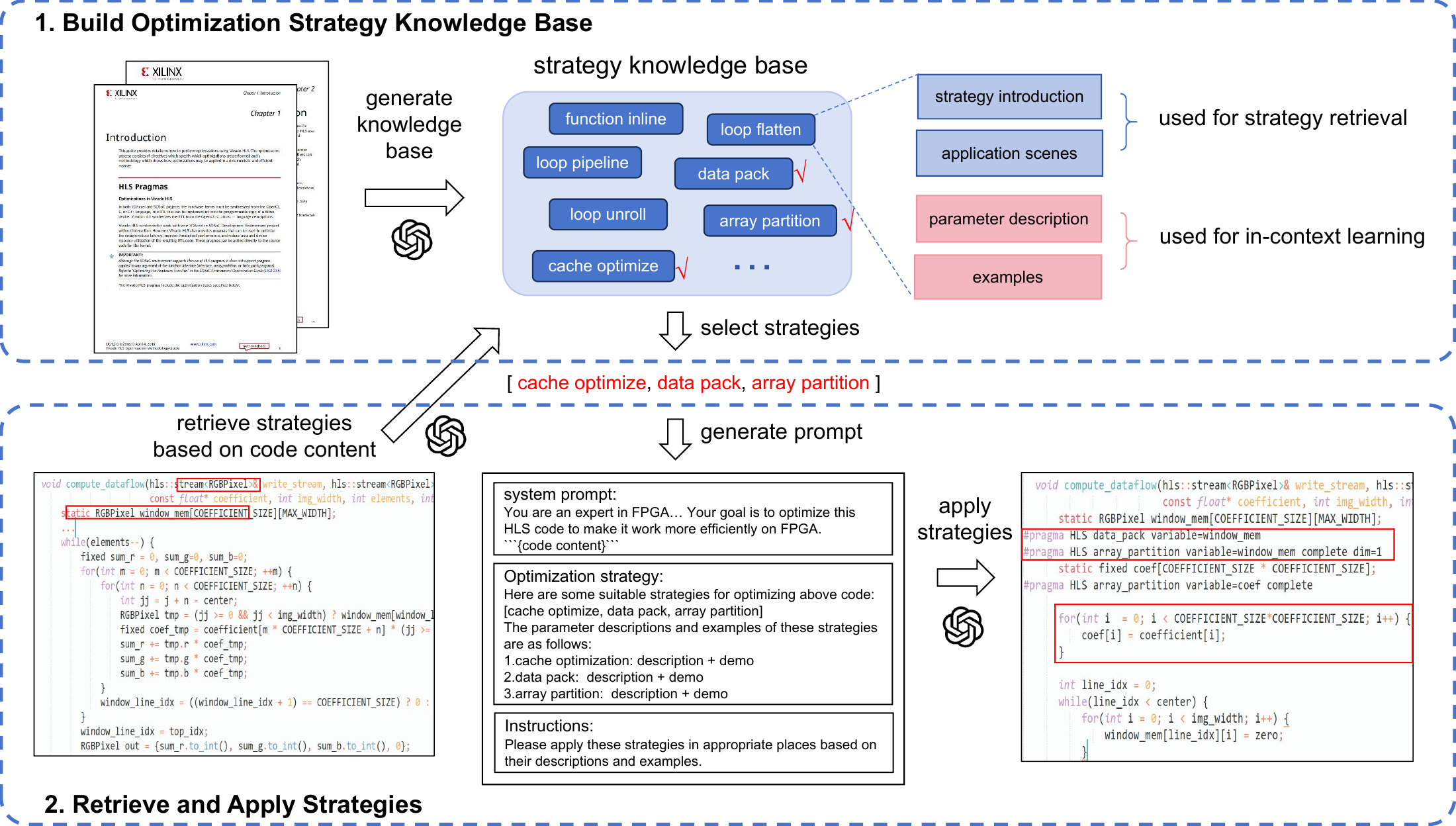}
    \caption{Automatic Optimization Strategies Learning and Application}
    \label{fig:stage-opt}
\end{figure*} 

The workflow of HLSPilot's RAG-like automatic optimization strategy learning is illustrated in Fig. \ref{fig:stage-opt}. 
It uses the Xilinx HLS official guide documentation as input and extracts structured pragma optimization information from the documents. As shown in fig. \ref{fig:pragma}, the structured information consists of four parts: (1) a brief introduction to the optimization strategy; (2) applicable optimization scenarios; (3) parameter descriptions; (4) optimization examples. The introduction to the optimization strategy and the information on applicable scenarios are primarily used to assist in retrieving and matching the optimization strategy with the code, thus these information is kept concise and general to enhance retrieval performance. Upon retrieving a suitable optimization strategy, the strategy's parameter description information and optimization example information are integrated into the prompt, utilizing the LLM's in-context learning capabilities to generate optimized code. 

\begin{figure}[htp]
    \centering
    \includegraphics[width=0.48\textwidth]{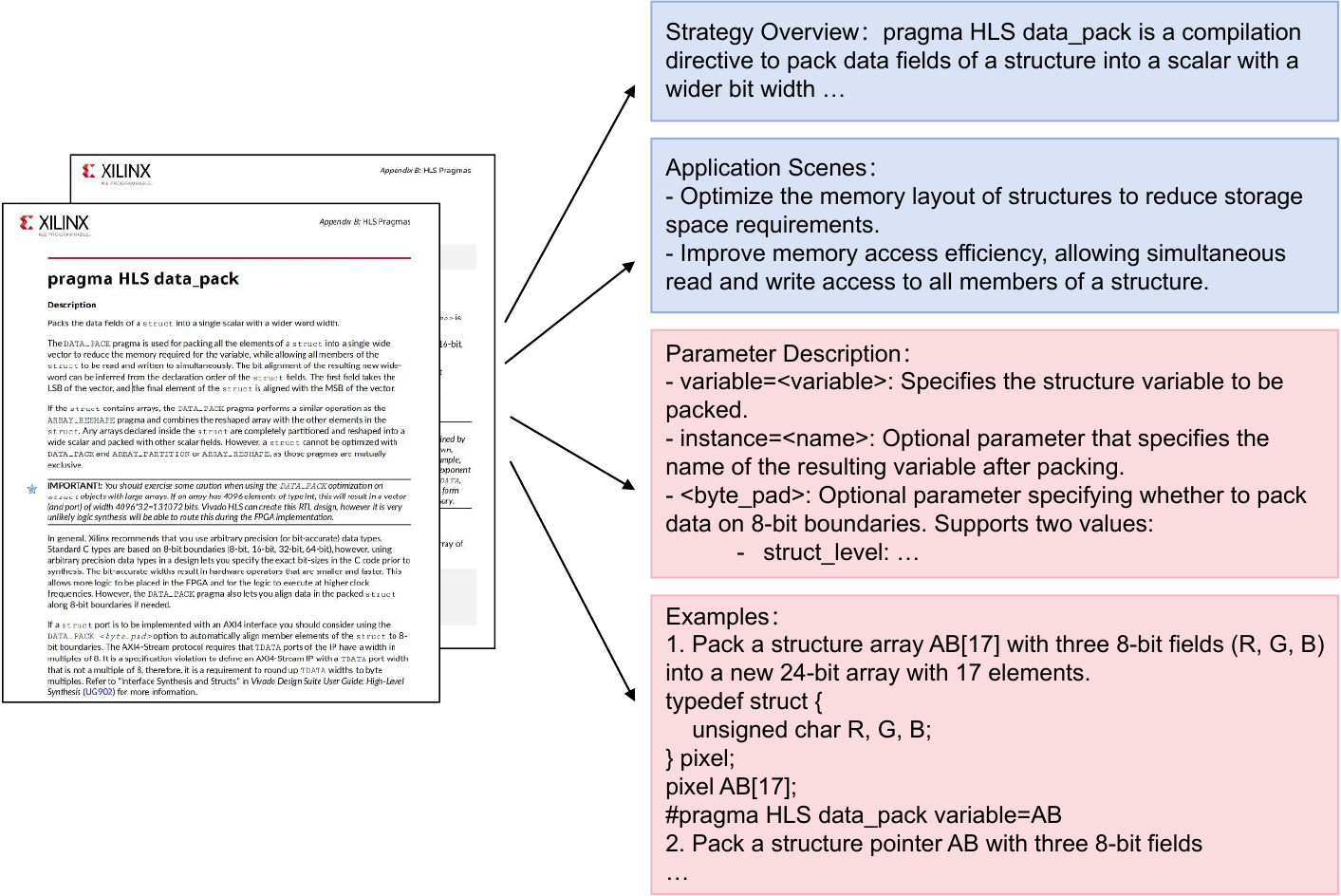}
    \caption{Structured information extracted by HLSPilot. The optimization strategy from documents is summarized into four parts: (1) strategy overview and (2) applicable scenarios for strategy retrieval; (3) parameter description and (4) examples for generating optimization prompt}
    \label{fig:pragma}
    \vspace{-1.2em}
\end{figure} 



\section{Experiment}
\subsection{Experiment Setting}
In this section, we demonstrate the effectiveness of HLSPilot framework for automatically generating and optimizing hardware accelerator based on HLS. We utilize GPT-4 \cite{GPT4} as the default LLM to accomplish tasks such as HLS code analysis and optimization within the workflow. For accelerator deployment and evaluation, we adopt the Vitis HLS design flow, using the Xilinx Alveo U280 data center accelerator card. For design space exploration, we utilizes GenHLSOptimizer \cite{DSEtool} to tune the parameters.

\subsection{Benchmark Introduction}
Currently most HLS benchmark suites \cite{hara2008chstone, schafer2014s2cbench, zhou2018rosetta} still face sevaral limitations. Firstly, many benchmarks are only comprised of some textbook-style function kernels, failing to fully implement the complexity of real-world applications. Thus evaluations on these benchmarks lack practical value. 
Secondly, most HLS benchmark suites only include optimized HLS designs, lacking corresponding unoptimized versions, which is unfriendly for evaluating the effectiveness of HLS optimization strategies. 

To address these issues and accurately evaluate the performance of the accelerators generated by our HLSPilot, we designed a benchmark suite that considers both the complexity of the designs and the convenience of comparing optimization effects. This benchmark suite consists of two parts: modified Rosetta benchmarks \cite{zhou2018rosetta} and a set of manually collected benchmarks. 
The Rosetta benchmarks comprise a series of complex real-world applications such as 3D rendering, digit recognition, and spam filtering. Each application has both a software implementation and a corresponding HLS implementation. The original Rosetta benchmarks were implemented using SDSoC. We have these designs ported to Vitis and proposed corresponding unoptimized HLS designs without any optimization strategies based on the software implementations of the applications. 
Additionally, as a supplement, we collected and implemented several other classic algorithm applications. Similarly, these applications also include unoptimized versions. 

\subsection{Experiment Results and Analysis}
\textbf{Experiment results}. Table \ref{table:runtime} shows the runtime of original unoptimized design, manually optimized design, HLSPilot-generated design, and HLSPilot-generated design with DSE for each application in the benchmarks. 
The results indicate a significant improvement in performance compared to the unoptimized design when utilizing HLSPilot-generated designs.
Overall, HLSPilot-generated designs achieve comparable performance to those manually optimized by human experts, while greatly reducing labor costs. With the utilization of DSE tools, some HLSPilot-generated designs can even outperform human designs. 

\begin{table}[h]
    \caption{Benchmark runtime(ms) on Xilinx Alveo U280}
        \label{table:runtime}
    \resizebox{0.5\textwidth}{!}{
    \renewcommand\arraystretch{1.4}
    \begin{tabular}{ccccc}
    \hline
    \textbf{Application}       & \textbf{original} & \textbf{handcrafted} & \textbf{HLSPilot} & \textbf{HLSPilot + DSE}         \\ \hline
    Fir               & 0.413    & 0.279      & \textbf{0.245}    & \textbf{0.227}         \\
    Merge Sort        & 786.618  & 54.878     & \textbf{47.580}   & \textbf{47.460}        \\
    BFS               & 5018.551 & \textbf{3973.645}   & 4184.273 & \textbf{3830.421}      \\
    PageRank          & 1862.214 & 1254.833   & \textbf{1114.991} & \textbf{1050.617}      \\
    3D Rendering      & 9.177    & \textbf{4.918}      & 5.375    & \textbf{5.146}         \\
    Digit Recognition & 9917.663 & \textbf{9.892}      & 78.837   & \textbf{52.832}        \\
    Face Detection    & 83.752   & \textbf{55.909}     & 64.372   & \textbf{59.138}        \\
    Optical Flow      & 101.313  & \textbf{54.084}     & 71.932   & \textbf{63.184}        \\
    Spam Filter       & 9278.917 & \textbf{37.346}     & 8013.913 & \textbf{7519.317}      \\ \hline
    \end{tabular}
    }
\end{table}

\textbf{Analysis on the results}. Table \ref{table:strategy} shows the major optimization strategies adopted by human expert's designs and HLSPilot-generated designs respectively.
It can be noted that HLSPilot has selected appropriate optimization strategies for different applications, basically covering the optimization selected by human expert. 
The performance gap between HLSPilot and human expert mainly comes from the specific implementation methods of optimization. For example, for dataflow pipelining optimization, there are various ways to split the same kernel. The rich experience of human experts may lead to more reasonable task partitioning. 
In addition, LLM struggles to implement optimizations tailored to specific scenes. For instance, in the spam filter application, achieving LUT optimization for sigmoid function requires sampling the function and generating a specific lookup table, while also considering issues such as quantization precision, which is difficult for LLM to implement.

\begin{table}[h]
    \caption{Major optimization strategies used in handcrafted design and HLSPilot-generated design}
        \label{table:strategy}
    \resizebox{0.5\textwidth}{!}{
    \renewcommand\arraystretch{1.2}
    \begin{tabular}{c|l|l}
        \hline
        \textbf{Application} &
          \textbf{manual} &
          \textbf{HLSPilot} \\ \hline
        Fir &
          \begin{tabular}[c]{@{}l@{}}Loop unrolling\\ Loop pipelining\end{tabular} &
          \begin{tabular}[c]{@{}l@{}}Loop unrolling\\ Loop pipeling\\ Memory optimization\end{tabular} \\ \hline
        Merge Sort &
          \begin{tabular}[c]{@{}l@{}}Dataflow pipelining\\ Memory Optimization\\ Loop unrolling\end{tabular} &
          \begin{tabular}[c]{@{}l@{}}Dataflow pipelining\\ Memory Optimization\\ Loop unrolling\end{tabular} \\ \hline
        BFS &
          \begin{tabular}[c]{@{}l@{}}Dataflow pipelining\\ Memory optimization\end{tabular} &
          \begin{tabular}[c]{@{}l@{}}Dataflow pipelining\\ Memory Optimization\end{tabular} \\ \hline
        PageRank &
          \begin{tabular}[c]{@{}l@{}}Dataflow pipelining\\ Memory optimization\end{tabular} &
          \begin{tabular}[c]{@{}l@{}}Dataflow pipelining\\ Memory optimization\end{tabular} \\ \hline
        3D Rendering &
          \begin{tabular}[c]{@{}l@{}}Dataflow pipelining\\ Communication optimization\end{tabular} &
          \begin{tabular}[c]{@{}l@{}}Dataflow pipelining\\ Communication optimization\\ Memory optimization\end{tabular} \\ \hline
        Digit Recognition &
          \begin{tabular}[c]{@{}l@{}}Loop unrolling\\ Loop pipelining\end{tabular} &
          \begin{tabular}[c]{@{}l@{}}Dataflow pipelining\\ Loop unrolling\\ Loop pipelining\\ Datatype optimization\end{tabular} \\ \hline
        Face Detection &
          \begin{tabular}[c]{@{}l@{}}Memory optimization\\ Datatype optimization\end{tabular} &
          \begin{tabular}[c]{@{}l@{}}Dataflow pipelining\\ Memory optimization\end{tabular} \\ \hline
        Optical Flow &
          \begin{tabular}[c]{@{}l@{}}Dataflow pipelining\\ Memory optimization\\ Communication optimization\end{tabular} &
          \begin{tabular}[c]{@{}l@{}}Dataflow pipelining\\ Memory optimization\\ Datatype optimization\\ Loop pipelining\end{tabular} \\ \hline
        Spam Filter &
          \begin{tabular}[c]{@{}l@{}}Dataflow pipelining\\ Memory optimization\\ Communication optimization\\ LUT optimization\end{tabular} &
          \begin{tabular}[c]{@{}l@{}}Dataflow pipelining\\ Memory optimization\end{tabular} \\ \hline
        \end{tabular}
    }
    \vspace{-1.2em}
\end{table}


\subsection{Case Study}
To further verify the practicality of HLSPilot in real-world application, we selected the L-BFGS algorithm \cite{liu1989limited} and performed a complete hardware acceleration workflow for it using HLSPilot on the hybrid CPU-FPGA platform.

\textbf{Introduction to the L-BFGS algorithm}. L-BFGS algorithm is one of the commonly used algorithms in machine learning for solving unconstrained optimization problems. 
When solving gradient descent, L-BFGS algorithm approximates the inverse Hessian matrix using only a limited amount of past information from the gradients, greatly reducing the storage space of data.
However, due to its large number of iterations, the algorithm performs poorly on the CPU, typically taking several hours for each search process.

\textbf{Complete acceleration workflow of HLSPilot}. In this case, we wrote a C++ software code for L-BFGS algorithm as the input of HLSPilot. 
HLSPilot firstly ran the sequential C++ code of the algorithm on CPU and generated a profiling report using the gprof tool, which includes detailed function runtime and number of calls. According to HLSPilot's analysis, the cost\_calculate function in L-BFGS accounts for more than 99.1\% of the total runtime of the algorithm, which is the performance bottleneck of the program. Therefore, this part will be extracted as the kernel for hardware acceleration.

Next, HLSPilot performed the task pipelining on the kernel code, partitioning the cost calculation process into three tasks: cost and convolution calculation, reconstruction error gradient calculation, and gradient check. Subsequently, HLSPilot applied appropriate optimization strategies to each task. The major optimization strategies employed in this stage included local buffer optimization, loop unrolling, array partitioning, and others. Particularly, HLSPilot noticed that the cost calculation process involved a significant amount of floating-point computations. Therefore, it performed floating-point to fixed-point conversion on the code, further optimizing the computational performance of the kernel. Finally, HLSPilot determined the pragma parameters through DSE tools.

\textbf{Acceleration result}. We evaluated the cost calculation runtime and algorithm's total runtime on both CPU and CPU-FPGA platforms, as shown in Table \ref{table:case-study-result}. L-BFGS-CPU represents the algorithm program running on the CPU, while HLSPilot-FP and HLSPilot-FXP respectively represent the floating-point and fixed-point designs generated by HLSPilot. Overall, HLSPilot's floating-point design and fixed-point design have accelerated the end-to-end runtime by 7.79 times and 11.93 times, respectively. Notably, for the cost calculation, HLSPilot can accelerate it by more than 500 times, which fully demonstrates the effectiveness of HLSPilot's acceleration.

\begin{table}[h]
    \vspace{-1em}
    \caption{Acceleration result on L-BFGS algorithm}
    \label{table:case-study-result}
    \vspace{-0.5em}
    \resizebox{0.5\textwidth}{!}{
    \renewcommand\arraystretch{1.2}
        \begin{tabular}{ccccc}
        \hline
        \textbf{Design} &
          \textbf{\begin{tabular}[c]{@{}c@{}}CostCalc.\\ Runtime(s)\end{tabular}} &
          \textbf{\begin{tabular}[c]{@{}c@{}}Total\\ Runtime(s)\end{tabular}} &
          \textbf{\begin{tabular}[c]{@{}c@{}}CostCalc.\\ Speedup\end{tabular}} &
          \textbf{\begin{tabular}[c]{@{}c@{}}End-to-end\\ Speedup\end{tabular}} \\ \hline
        CPU          & 18237 & 18390 & -                & -               \\
        HLSPilot-FP  & 855   & 2365  & 21.33x           & 7.78x           \\
        HLSPilot-FXP & 31    & 1541  & \textbf{588.29x} & \textbf{11.93x} \\ \hline
        \end{tabular}
    }
\end{table}

Table \ref{table:case-study-result2} shows the resource overhead and runtime of the cost calculation kernel in L-BFGS. Runtime in table \ref{table:case-study-result2} represents the time taken to execute one instance of the cost calculation. It is evident that HLSPilot can effectively optimize the performance bottlenecks of the algorithm, significantly enhancing performance.

\begin{table}[h]
    \vspace{-1em}
    \caption{CostCalc. kernel resource overhead and runtime}
    \label{table:case-study-result2}
    \vspace{-0.5em}
    \resizebox{0.5\textwidth}{!}{
    \renewcommand\arraystretch{1.3}
    \begin{tabular}{cccccc}
    \hline
    \textbf{Kernels} & \textbf{\#LUTs} & \textbf{\#FFs} & \textbf{\#BRAMs} & \textbf{\#DSPs} & \textbf{Runtime(ms)} \\ \hline
    CPU        & -      & -      & -   & -   & 38529.08 \\
    kernel-FP  & 54970  & 66459  & 46  & 107 & 1680.84  \\
    kernel-FXP & 188294 & 245018 & 270 & 624 & 60.9811  \\ \hline
    \end{tabular}
    }
    \vspace{-1em}
\end{table}

\section{Conclusion}

In this paper, we have introduced HLSPilot, the first LLM-driven HLS framework to automate the generation of hardware accelerators on CPU-FPGA platform. HLSPilot focuses on the transformation between sequential C/C++ code and optimized HLS code, which greatly reducing the semantic gap between design intent and hardware code. Additionally, the integration of profiling tools and DSE tools enables automatic hardware/software partition and pragma tuning. Through the combined efforts of various modules driven by LLM, HLSPilot automatically generates high-performance hardware accelerators.
The kernel optimization experiment results on the benchmark fully demonstrate the potential of HLSPilot, showing its ability to achieve comparable, and in some cases superior, performance relative to manually designed FPGA kernel. In addition, we also performed a complete hardware acceleration workflow for a real-world algorithm, achieving 11.93x speedup on the hybrid CPU-FPGA platform. These results highlight the significant effects of LLM, suggesting a promising future for LLM-assisted methodology in hardware design.


\newpage
\bibliographystyle{IEEEtran}
\bibliography{hlspilotreferences}

\begin{thebibliography}{10}
\providecommand{\url}[1]{#1}
\csname url@samestyle\endcsname
\providecommand{\newblock}{\relax}
\providecommand{\bibinfo}[2]{#2}
\providecommand{\BIBentrySTDinterwordspacing}{\spaceskip=0pt\relax}
\providecommand{\BIBentryALTinterwordstretchfactor}{4}
\providecommand{\BIBentryALTinterwordspacing}{\spaceskip=\fontdimen2\font plus
\BIBentryALTinterwordstretchfactor\fontdimen3\font minus \fontdimen4\font\relax}
\providecommand{\BIBforeignlanguage}[2]{{%
\expandafter\ifx\csname l@#1\endcsname\relax
\typeout{** WARNING: IEEEtran.bst: No hyphenation pattern has been}%
\typeout{** loaded for the language `#1'. Using the pattern for}%
\typeout{** the default language instead.}%
\else
\language=\csname l@#1\endcsname
\fi
#2}}
\providecommand{\BIBdecl}{\relax}
\BIBdecl

\bibitem{martin2009high}
G.~Martin and G.~Smith, ``High-level synthesis: Past, present, and future,'' \emph{IEEE Design \& Test of Computers}, vol.~26, no.~4, pp. 18--25, 2009.

\bibitem{liu2019obfs}
C.~Liu, X.~Chen, B.~He, X.~Liao, Y.~Wang, and L.~Zhang, ``Obfs: Opencl based bfs optimizations on software programmable fpgas,'' in \emph{2019 International Conference on Field-Programmable Technology (ICFPT)}.\hskip 1em plus 0.5em minus 0.4em\relax IEEE, 2019, pp. 315--318.

\bibitem{zhang2024graphitron}
X.~Zhang, Z.~Feng, S.~Liang, X.~Chen, C.~Liu, H.~Li, and X.~Li, ``Graphitron: A domain specific language for fpga-based graph processing accelerator generation,'' \emph{arXiv preprint arXiv:2407.12575}, 2024.

\bibitem{lahti2018we}
S.~Lahti, P.~Sj{\"o}vall, J.~Vanne, and T.~D. H{\"a}m{\"a}l{\"a}inen, ``Are we there yet? a study on the state of high-level synthesis,'' \emph{IEEE Transactions on Computer-Aided Design of Integrated Circuits and Systems}, vol.~38, no.~5, pp. 898--911, 2018.

\bibitem{schafer2019high}
B.~C. Schafer and Z.~Wang, ``High-level synthesis design space exploration: Past, present, and future,'' \emph{IEEE Transactions on Computer-Aided Design of Integrated Circuits and Systems}, vol.~39, no.~10, pp. 2628--2639, 2019.

\bibitem{zhao2019performance}
J.~Zhao, L.~Feng, S.~Sinha, W.~Zhang, Y.~Liang, and B.~He, ``Performance modeling and directives optimization for high-level synthesis on fpga,'' \emph{IEEE Transactions on Computer-Aided Design of Integrated Circuits and Systems}, vol.~39, no.~7, pp. 1428--1441, 2019.

\bibitem{liu2015quickdough}
C.~Liu, H.-C. Ng, and H.~K.-H. So, ``Quickdough: A rapid fpga loop accelerator design framework using soft cgra overlay,'' in \emph{2015 International Conference on Field Programmable Technology (FPT)}.\hskip 1em plus 0.5em minus 0.4em\relax IEEE, 2015, pp. 56--63.

\bibitem{sohrabizadeh2022autodse}
A.~Sohrabizadeh, C.~H. Yu, M.~Gao, and J.~Cong, ``Autodse: Enabling software programmers to design efficient fpga accelerators,'' \emph{ACM Transactions on Design Automation of Electronic Systems (TODAES)}, vol.~27, no.~4, pp. 1--27, 2022.

\bibitem{choi2018hls}
Y.-k. Choi and J.~Cong, ``Hls-based optimization and design space exploration for applications with variable loop bounds,'' in \emph{2018 IEEE/ACM International Conference on Computer-Aided Design (ICCAD)}.\hskip 1em plus 0.5em minus 0.4em\relax IEEE, 2018, pp. 1--8.

\bibitem{zhong2017design}
G.~Zhong, A.~Prakash, S.~Wang, Y.~Liang, T.~Mitra, and S.~Niar, ``Design space exploration of fpga-based accelerators with multi-level parallelism,'' in \emph{Design, Automation \& Test in Europe Conference \& Exhibition (DATE), 2017}.\hskip 1em plus 0.5em minus 0.4em\relax IEEE, 2017, pp. 1141--1146.

\bibitem{ferretti2022graph}
L.~Ferretti, A.~Cini, G.~Zacharopoulos, C.~Alippi, and L.~Pozzi, ``Graph neural networks for high-level synthesis design space exploration,'' \emph{ACM Transactions on Design Automation of Electronic Systems}, vol.~28, no.~2, pp. 1--20, 2022.

\bibitem{luo2023deepburning}
E.~Luo, H.~Huang, C.~Liu, G.~Li, B.~Yang, Y.~Wang, H.~Li, and X.~Li, ``Deepburning-mixq: An open source mixed-precision neural network accelerator design framework for fpgas,'' in \emph{2023 IEEE/ACM International Conference on Computer Aided Design (ICCAD)}.\hskip 1em plus 0.5em minus 0.4em\relax IEEE, 2023, pp. 1--9.

\bibitem{chen2021thundergp}
X.~Chen, H.~Tan, Y.~Chen, B.~He, W.-F. Wong, and D.~Chen, ``Thundergp: Hls-based graph processing framework on fpgas,'' in \emph{The 2021 ACM/SIGDA International Symposium on Field-Programmable Gate Arrays}, 2021, pp. 69--80.

\bibitem{liang2020deepburning}
S.~Liang, C.~Liu, Y.~Wang, H.~Li, and X.~Li, ``Deepburning-gl: an automated framework for generating graph neural network accelerators,'' in \emph{Proceedings of the 39th International Conference on Computer-Aided Design}, 2020, pp. 1--9.

\bibitem{fu2023gpt4aigchip}
Y.~Fu, Y.~Zhang, Z.~Yu, S.~Li, Z.~Ye, C.~Li, C.~Wan, and Y.~C. Lin, ``Gpt4aigchip: Towards next-generation ai accelerator design automation via large language models,'' in \emph{2023 IEEE/ACM International Conference on Computer Aided Design (ICCAD)}.\hskip 1em plus 0.5em minus 0.4em\relax IEEE, 2023, pp. 1--9.

\bibitem{chang2023chipgpt}
K.~Chang, Y.~Wang, H.~Ren, M.~Wang, S.~Liang, Y.~Han, H.~Li, and X.~Li, ``Chipgpt: How far are we from natural language hardware design,'' \emph{arXiv preprint arXiv:2305.14019}, 2023.

\bibitem{liu2023chipnemo}
M.~Liu, T.-D. Ene, R.~Kirby, C.~Cheng, N.~Pinckney, R.~Liang, J.~Alben, H.~Anand, S.~Banerjee, I.~Bayraktaroglu \emph{et~al.}, ``Chipnemo: Domain-adapted llms for chip design,'' \emph{arXiv preprint arXiv:2311.00176}, 2023.

\bibitem{thakur2023benchmarking}
S.~Thakur, B.~Ahmad, Z.~Fan, H.~Pearce, B.~Tan, R.~Karri, B.~Dolan-Gavitt, and S.~Garg, ``Benchmarking large language models for automated verilog rtl code generation,'' in \emph{2023 Design, Automation \& Test in Europe Conference \& Exhibition (DATE)}.\hskip 1em plus 0.5em minus 0.4em\relax IEEE, 2023, pp. 1--6.

\bibitem{liu2023verilogeval}
M.~Liu, N.~Pinckney, B.~Khailany, and H.~Ren, ``Verilogeval: Evaluating large language models for verilog code generation,'' in \emph{2023 IEEE/ACM International Conference on Computer Aided Design (ICCAD)}.\hskip 1em plus 0.5em minus 0.4em\relax IEEE, 2023, pp. 1--8.

\bibitem{tsai2023rtlfixer}
Y.~Tsai, M.~Liu, and H.~Ren, ``Rtlfixer: Automatically fixing rtl syntax errors with large language models,'' \emph{arXiv preprint arXiv:2311.16543}, 2023.

\bibitem{liu2023rtlcoder}
S.~Liu, W.~Fang, Y.~Lu, Q.~Zhang, H.~Zhang, and Z.~Xie, ``Rtlcoder: Outperforming gpt-3.5 in design rtl generation with our open-source dataset and lightweight solution,'' \emph{arXiv preprint arXiv:2312.08617}, 2023.

\bibitem{thakur2023autochip}
S.~Thakur, J.~Blocklove, H.~Pearce, B.~Tan, S.~Garg, and R.~Karri, ``Autochip: Automating hdl generation using llm feedback,'' \emph{arXiv preprint arXiv:2311.04887}, 2023.

\bibitem{lu2024rtllm}
Y.~Lu, S.~Liu, Q.~Zhang, and Z.~Xie, ``Rtllm: An open-source benchmark for design rtl generation with large language model,'' in \emph{2024 29th Asia and South Pacific Design Automation Conference (ASP-DAC)}.\hskip 1em plus 0.5em minus 0.4em\relax IEEE, 2024, pp. 722--727.

\bibitem{blocklove2023chip}
J.~Blocklove, S.~Garg, R.~Karri, and H.~Pearce, ``Chip-chat: Challenges and opportunities in conversational hardware design,'' in \emph{2023 ACM/IEEE 5th Workshop on Machine Learning for CAD (MLCAD)}.\hskip 1em plus 0.5em minus 0.4em\relax IEEE, 2023, pp. 1--6.

\bibitem{jiang2024iicpilot}
Z.~Jiang, Q.~Zhang, C.~Liu, H.~Li, and X.~Li, ``Iicpilot: An intelligent integrated circuit backend design framework using open eda,'' \emph{arXiv preprint arXiv:2407.12576}, 2024.

\bibitem{nijkamp2022codegen}
E.~Nijkamp, B.~Pang, H.~Hayashi, L.~Tu, H.~Wang, Y.~Zhou, S.~Savarese, and C.~Xiong, ``Codegen: An open large language model for code with multi-turn program synthesis,'' \emph{arXiv preprint arXiv:2203.13474}, 2022.

\bibitem{chen2021evaluating}
M.~Chen, J.~Tworek, H.~Jun, Q.~Yuan, H.~P. d.~O. Pinto, J.~Kaplan, H.~Edwards, Y.~Burda, N.~Joseph, G.~Brockman \emph{et~al.}, ``Evaluating large language models trained on code,'' \emph{arXiv preprint arXiv:2107.03374}, 2021.

\bibitem{wang2021codet5}
Y.~Wang, W.~Wang, S.~Joty, and S.~C. Hoi, ``Codet5: Identifier-aware unified pre-trained encoder-decoder models for code understanding and generation,'' \emph{arXiv preprint arXiv:2109.00859}, 2021.

\bibitem{wang2024grammar}
B.~Wang, Z.~Wang, X.~Wang, Y.~Cao, R.~A~Saurous, and Y.~Kim, ``Grammar prompting for domain-specific language generation with large language models,'' \emph{Advances in Neural Information Processing Systems}, vol.~36, 2024.

\bibitem{schick2024toolformer}
T.~Schick, J.~Dwivedi-Yu, R.~Dess{\`\i}, R.~Raileanu, M.~Lomeli, E.~Hambro, L.~Zettlemoyer, N.~Cancedda, and T.~Scialom, ``Toolformer: Language models can teach themselves to use tools,'' \emph{Advances in Neural Information Processing Systems}, vol.~36, 2024.

\bibitem{DSEtool}
\BIBentryALTinterwordspacing
aferikoglou, ``Genhlsoptimizer,'' 2022. [Online]. Available: \url{https://github.com/aferikoglou/GenHLSOptimizer}
\BIBentrySTDinterwordspacing

\bibitem{XilinxUG902}
\BIBentryALTinterwordspacing
\vspace{0mm}Xilinx, ``Vivado design suite user guide: High-level synthesis (ug902),'' 2020. [Online]. Available: \url{https://docs.amd.com/v/u/en-US/ug902-vivado-high-level-synthesis}
\BIBentrySTDinterwordspacing

\bibitem{XilinxUG1270}
\BIBentryALTinterwordspacing
Xilinx, ``Vivado hls optimization methodology guide (ug1270),'' 2018. [Online]. Available: \url{https://docs.amd.com/v/u/en-US/ug1270-vivado-hls-opt-methodology-guide}
\BIBentrySTDinterwordspacing

\bibitem{XilinxUG1399}
\BIBentryALTinterwordspacing
\vspace{0mm}Xilinx, ``Vitis high-level synthesis user guide (ug1399),'' 2023. [Online]. Available: \url{https://docs.amd.com/r/en-US/ug1399-vitis-hls/Navigating-Content-by-Design-Process}
\BIBentrySTDinterwordspacing

\bibitem{GPT4}
\BIBentryALTinterwordspacing
OpenAI, ``Gpt-4,'' 2023. [Online]. Available: \url{https://platform.openai.com/docs/models/gpt-4-and-gpt-4-turbo}
\BIBentrySTDinterwordspacing

\bibitem{hara2008chstone}
Y.~Hara, H.~Tomiyama, S.~Honda, H.~Takada, and K.~Ishii, ``Chstone: A benchmark program suite for practical c-based high-level synthesis,'' in \emph{2008 IEEE International Symposium on Circuits and Systems (ISCAS)}.\hskip 1em plus 0.5em minus 0.4em\relax IEEE, 2008, pp. 1192--1195.

\bibitem{schafer2014s2cbench}
B.~C. Schafer and A.~Mahapatra, ``S2cbench: Synthesizable systemc benchmark suite for high-level synthesis,'' \emph{IEEE Embedded Systems Letters}, vol.~6, no.~3, pp. 53--56, 2014.

\bibitem{zhou2018rosetta}
Y.~Zhou, U.~Gupta, S.~Dai, R.~Zhao, N.~Srivastava, H.~Jin, J.~Featherston, Y.-H. Lai, G.~Liu, G.~A. Velasquez \emph{et~al.}, ``Rosetta: A realistic high-level synthesis benchmark suite for software programmable fpgas,'' in \emph{Proceedings of the 2018 ACM/SIGDA International Symposium on Field-Programmable Gate Arrays}, 2018, pp. 269--278.

\bibitem{liu1989limited}
D.~C. Liu and J.~Nocedal, ``On the limited memory bfgs method for large scale optimization,'' \emph{Mathematical programming}, vol.~45, no.~1, pp. 503--528, 1989.

\end{thebibliography}

%

\end{document}